\documentclass{article}

\usepackage[backend=biber,style = phys,eprint=true]{biblatex}
\usepackage{amsmath,amssymb,amsfonts}
\usepackage{graphicx}
\usepackage{authblk}
\usepackage{hyperref}
\usepackage{cleveref}
\usepackage{srcltx}
\usepackage{bm}
\usepackage{mathdots}
\usepackage{color}
\usepackage{dsfont}
\usepackage{slashed}
\usepackage{physics}
\usepackage{cancel}
\usepackage{braket}
\usepackage{comment}

\title{Exploration to early universe by Josephson Junction Switching Current Detector}

\author{Dan Kondo
\footnote{\href{mailto:dan.kondo@ipmu.jp}{dan.kondo@ipmu.jp} } }
\affil{Kavli IPMU (WPI), UTIAS, The University of Tokyo, Kashiwa, Chiba 277-8583, Japan}

\date{}

\begin{document}

\maketitle

\begin{abstract}
    In this paper, we propose a method to probe a Stochastic Gravitational Wave Background (SGWB) with Josephson Junction Switching Current Detector (JJSCD). The sensitivity for the shear can reach $h\simeq 10^{-19}$ realistically, $10^{-21}$ in the near future, $10^{-24}$ optimistically. If we utilize the enhancement factor from the ratio of the frequency, it is possible to reach further below the Big Bang Nucleosynthesis (BBN) bound. It will be interesting if we can access the region to discover a footprint of new physics.
\end{abstract}

\section{Introduction}
Ultra High frequency Gravitational Waves (UHF-GWs) can be a frontier for exploring new physics in the next generation. On one hand, there are no known source of gravitational wave with frequency higher than kHz from astrophysical side \cite{Aggarwal:2020olq}. On the other hand, there are proposal for the possible sources of UHF-GWs such as Primordial Black Hole (PBH) \cite{Carr:2020gox,Escriva:2022duf}, inflation\cite{Barnaby:2010vf,Domcke:2016bkh,Garcia-Bellido:2016dkw,Bartolo:2015qvr,Graef:2015ova,Cannone:2014uqa,Ricciardone:2016lym}, reheating\cite{Figueroa:2017vfa,Figueroa:2014aya,Enqvist:2012im,Benakli:2018xdi,Adshead:2018doq,Adshead:2019lbr,Lozanov:2019jxc}, oscillon\cite{Amin:2010dc,Amin:2010jq,Gleiser:2008ty,Lozanov:2017hjm,Lozanov:2019ylm,Lozanov:2023aez,Lozanov:2023knf} and topological defects\cite{Figueroa:2012kw,Auclair:2019wcv,Dror:2019syi,Dunsky:2021tih}. Detection of such a GW can be in the frontier for the next generation of physics. There are proposals for the hunting gravitational waves. Thanks to the fact that the frequency of interest is close to the typical mass of QCD axion \cite{ParticleDataGroup:2022pth,AxionLimits,Chigusa:2023szl}, they are mainly from applying axion detection experimental proposal to the GW detection such as haloscope, magnon, Rydberg atom sensor \cite{Ringwald:2020ist,Domcke:2022rgu,Domcke:2023bat,Domcke:2023qle,Ito:2020wxi,Ito:2022rxn,Kanno:2023whr,Fan2015AtomBR,schmidt2023rydberg,Kanno:2023whr}.
Other than axion method, we can consider a way to utilize the property of gravity. One of such way is to use the preciseness of atomic clock \cite{Bringmann:2023gba}. In this paper, they comment that we need a way to probe the Stochastic Gravittaional Wave Background (SGWB). In this paper, we propose a way to hunt the possibility with application of Josephson Junction Switching Current Detector (JJSCD) \cite{krasnov2024resonant}. The punchline to obtain the small signal is to apply for the technique of heterodyne(homodyne) detection \cite{Protopopov2009}. The idea is simple. First thing to do is to obtain a modulation signal from the combination of EM background wave and SGWB. Second, we make the modulation signal pass to drop the other contributions. Finally, we will stimulate the JJSCD to count the signal. The concept of the JJSCD is similar to transition edge sensor (TES). The benefit of UHFGW search is that we can probe with compact size apparatus because of the shortness of the wave length.

The organization of this paper is in the following way. In \cref{sec:EWandGW}, we will see a relation between electro-magnetic (EM) wave and gravitational wave. In \cref{sec:detectionscheme}, we will see a way 
to pick up the modulation signal. In \cref{sec:JJSCD}, we will introduce a concept of Josephson junction switching current detector. In \cref{sec:sensitivityestimate}, we estimate the sensitivity of the gravitational wave shear in three cases, non-resonant (realistic), resonant (near future), and optimistic. 

\section{EM and GW}\label{sec:EWandGW}
In this section, we will see the relation between the electromagnetic wave and gravitational wave based on \cite{Bringmann:2023gba}. There are mainly two way to see the GW effects on the propagating EM waves. One is to see the shift of the frequency, the other is to see the modulation. In the former case, we can see the EM waves are red-shifted in the presence of gravity at the level of $\mathcal{O}(h)$. We wrote about it in \cref{sec:Photonfrequencyshift}. In the latter case, we can see the modulated EM waves. EM contains tiny components mixed with GW with shifted frequency. We will explore in the following section in \cref{sec:EMprop}.

\subsection{Propagation of EM wave in curved spacetime}\label{sec:EMprop}
Then Maxwell equation in perturbed curved spacetime is \cite{Landau1952ClassicalTO}
\begin{align}
     \partial_\mu F^{\mu\nu}&=-\partial_\mu \left[\frac{1}{2}hF^{\mu\nu}-h^{\mu\alpha}F_\alpha^\nu-F^\mu_\beta h^{\beta\nu}\right]
\end{align}

Here, we expand $F_{\mu\nu}=F^0_{\mu\nu}+\Delta F_{\mu\nu}$ as $\mathcal{O}(h^0)$ part and $\mathcal{O}(h)$ part. $F^0_{\mu\nu}$ is the solution of Maxwell equation in flat spacetime $\partial_\mu F^{\mu\nu}=0$.

To the level of $\mathcal{O}(h)$, the Maxwell equation is
\begin{align}\label{eq:MaxwellOh}
    \partial_\mu\Delta F^{\mu\nu}
    =-\partial_\mu \left[\frac{1}{2}h(F_0)^{\mu\nu}-h^{\mu\alpha}(F_0)_{\alpha}^\nu-(F_0)^\mu_{\beta} h^{\beta\nu}\right]=j_{\text{eff}}^\nu.
\end{align}
From this equation, we can read out that the presence of GW and background EM wave induce an EM wave with amplitude $\mathcal{O}(h)$ compared to original one.
In TT gauge, 
\begin{align}
    h^{TT}(x^\mu)=h_+ s^2_\theta \cos[\omega_g(t-c_\theta x-s_\theta z)+\varphi_0]
\end{align}

In transverse traceless gauge (TT gauge), a plame GW propagating in $x^3-x^1$ plane with angle $\theta$ can be written as \cite{Domcke:2024abs} 
\begin{align}
h^{TT}_{t\mu}&=h^{TT}_{\mu t}=0,\\
h^{TT}_{ij}&=(h_+e^+_{ij}+h_\times e^\times_{ij})\cos[\omega_g(x^0-\hat{\bm{k}}_g\cdot\bm{x})+\varphi_0], \\
\hat{\bm{k}}_g&= (c_\theta,0,s_\theta)^T,\\
e^+_{ij}&=
\begin{pmatrix}
    s_\theta^2&0&-s_\theta c_\theta\\
    0&-1&0\\
    -s_\theta c_\theta&0&c^2_\theta
\end{pmatrix},\
e^\times_{ij}=
\begin{pmatrix}
    0& c_\theta&0\\
    c_\theta&0&-c_\theta\\
    0&-c_\theta&0
\end{pmatrix}.
\end{align}

Let us consider the following EM wave 
\begin{align}
    \bm{A}(t,x)&=
    A_0\cos[\omega_\gamma(t-x)]\bm{e}_z+\bm{\Delta}\bm{A}(t,x)
\end{align}
with field strength for the flat part 
\begin{align}
    (F_0)_{tx}&=(F_0)_{ty}=(F_0)_{xy}=(F_0)_{yz}=0, \\
    (F_0)_{tz}&=
    -A_0\omega_\gamma \sin  
\omega_\gamma(t-x), \\
    (F_0)_{xz}&= A_0\omega_{\gamma}\sin\omega_\gamma(t-x).
\end{align}
With combination of $h^{TT}$, the right hand side of \ref{eq:MaxwellOh} is the linear combination of 
\begin{align}
    \cos_-&= \cos[(\omega_g-\omega_\gamma)t-(\omega_gc_\theta-\omega_\gamma)x],\\
    \cos_+&= \cos[(\omega_g+\omega_\gamma)t-(\omega_gc_\theta+\omega_\gamma)x].
\end{align}
We assume that $\Delta \bm{A}$ takes the following form
\begin{align}
    \Delta\bm{A}_i(t,\bm{x})&=
    a_i^-\cos_-+a_i^+\cos_+.
\end{align}
We can evaluate the both the left hand side and the right hand side of the equation \cref{eq:MaxwellOh} to obtain the EM amplitude of $\mathcal{O}(h)$. You can see the \cref{sec:Caldetail} for the calculation process.

\begin{align}
    a^{\pm}_x&= -\frac{1}{4}A_0h_+s_{2\theta}\frac{\omega_\gamma(\omega_\gamma\pm\omega_g(1-c_\theta))}{(\omega_\gamma\pm\omega_g)^2}
    \simeq -\frac{1}{4}A_0h_+s_{2\theta}+\mathcal{O}\left(\frac{\omega_g}{\omega_\gamma}\right),\\
    a^{\pm}_y&= \frac{\pm A_0h_\times\omega_\gamma\omega_g c_\theta}{2\{\omega_g^2(1+c_\theta)\pm 2\omega_g\omega_\gamma\}}
    \simeq \frac{1}{4}A_0h_\times c_\theta+\mathcal{O}\left(\frac{\omega_g}{\omega_\gamma}\right),\\
    a^\pm_z&= \frac{A_0h_+\omega_\gamma\{\omega_\gamma(1+c_\theta)\pm c_\theta^2\omega_g\} }{2\{\omega_g^2(1+c_\theta)\pm 2\omega_g\omega_\gamma\} }
    \simeq \pm\frac{1}{4}A_0h_+\frac{\omega_\gamma}{\omega_g}(1+c_\theta)+\mathcal{O}\left(\frac{\omega_g}{\omega_\gamma}\right).
\end{align}
The point here is that all of them are $\mathcal{O}(h)$ relative to background EM wave amplitude. This means that background EM wave and GW produce EM wave with amplitude $\mathcal{O}(h)$ compared to original amplitude with shifted frequency. Also, along the direction of input EM wave, there is an enhancement factor $\omega_\gamma/\omega_g$. It is more likely that we can hunt a signal from $h_+$ for high frequency input EM wave and relatively low frequency GW. Note that both $h_+$ and $h_\times$ arise in the direction perpendicular to the input EM wave. If there are chances to detect the signal in the future, by comparing the signal orthogonal to the input direction, we will be able to study the ratio of $h_+$ and $h_\times$, which can be a hint of early universe.

\section{Detection scheme}\label{sec:detectionscheme}
\begin{figure}
    \centering
    \includegraphics[width=10cm]{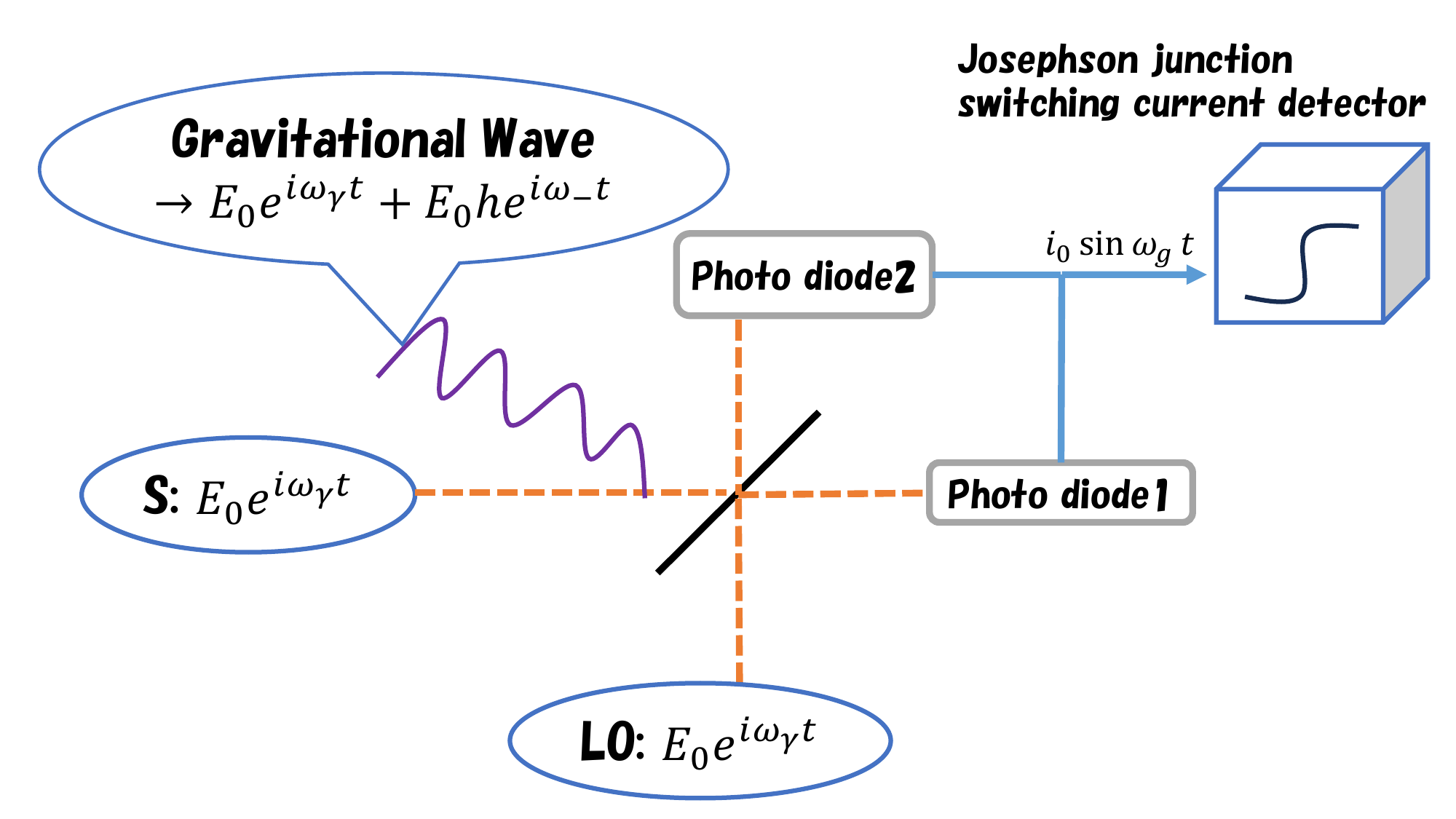}
    \caption{Schematic figure of the detection method. By injecting laser to the photo diode and take difference between two. If there is a gravitational wave background, background EM field and GW produce a small signal. There is a modulation signal from two input. The output from the photo diode produce a microwave current with frequency $\omega_g$. This small current causes a switching from superconducting state to normal state of Josephson junction.}
    \label{fig:schematichomodyne}
\end{figure}

We will describe how to detect the effect from GW. We can write the photon wave at the detector as \cite{Bringmann:2023gba}
\begin{align}
    &A(t,L)\nonumber\\
    =&A_\gamma \cos(\omega^s_\gamma t+\phi'_\gamma)+A_\gamma\frac{h_+}{4}\frac{\omega^s_\gamma}{\omega_g}\left[\cos[\omega^+(t-L)+\phi'_\gamma]-\cos[\omega^-(t-L)+\phi'_\gamma]\right] \nonumber\\
    +&A_\gamma\frac{h_+}{4}\frac{\omega^s_\gamma}{\omega_g}\left[-\cos[\omega^+t-\omega_\gamma L+\phi'_\gamma]+\cos[\omega^-t-\omega_\gamma L+\phi'_\gamma]\right]\nonumber\\
    -&A_\gamma\frac{h_+}{2}\frac{\omega^s_\gamma}{\omega_g}\left[\sin[\omega_gL]\sin[\omega_\gamma(t-L)+\phi'_\gamma]\right] \nonumber
\end{align}
This is linear combination of each basic waves, we will pick up a second term as representative wave. We write it as
\begin{align}
    A(t,L)&= A_\gamma\cos(\omega_\gamma t+\phi'_\gamma)
    -A_\gamma\frac{h_+}{4}\frac{\omega^s_\gamma}{\omega_g}[\cos[\omega^-(t-L)+\phi'_\gamma]]
\end{align}
The electric field an be obtained as
\begin{align}
E&= -\frac{\partial A(t,L)}{\partial t} \nonumber\\
&=A_\gamma\omega^s_\gamma\sin(\omega^s_\gamma t+\phi'_\gamma)
    -A_\gamma\omega^s_\gamma\frac{h_+}{4}\frac{\omega^-}{\omega_g}[\sin[\omega^-(t-L)+\phi'_\gamma]] \nonumber\\
    &= E_0\sin(\omega^s_\gamma t+\phi'_\gamma)
    -E_0h[\sin[\omega^-(t-L)+\phi'_\gamma]]
\end{align}
where we wrote $h=\frac{h_+}{4}\frac{\omega^-}{\omega_g}$. From this expression, we can see that if GW comes, we can see the modulation in the signal. To test the possibility, we propose the way to detect the effect shown in \ref{fig:schematichomodyne}.
The light source emit laser beam with frequency $\omega^s_\gamma$. The beam is split into half by half by $50\%$ beam splitter. The first beam arrives at detector1 as 
\begin{align}
    E_1&= E_0 \mathrm{e}^{i\omega^s_\gamma t+\phi'_\gamma}. 
\end{align}
The second wave affected by GW arrives at detector2 as
\begin{align}
    E_2&= 
    E_0\mathrm{e}^{\omega^s_\gamma t+\phi'_\gamma}
    +E_0h\mathrm{e}^{\omega^-(t-L)+\phi'_\gamma}
\end{align}
The intensity of the laser is Poynting vector $\bm{S}=\bm{E}\times \bm{H}$,
\begin{align}
    I_1&= \frac{c\epsilon_0}{2}|E_1|^2
    =\frac{c\epsilon_0}{2}E_0^2 \equiv P_0 \nonumber\\
    I_2&= \frac{c\epsilon_0}{2}|E_2|^2
    =\frac{c\epsilon_0}{2}E_0^2+c\epsilon_0E_0^2h\cos(\omega^-(t-L)-\omega^s_\gamma t)+\frac{c\epsilon_0}{2}E_0^2h^2\nonumber\\
&=P_0+ 2P_0 h\cos(\omega^-(t-L)-\omega^s_\gamma t)+P_0h^2    
\end{align}
Here, we use a low pass filter which shuts out the original laser with frequency $\omega_\gamma^s$. The last term is negligible considering that $h<10^{-20}$ from BBN bound \cite{Caprini:2018mtu,Cyburt:2015mya,Burns:2023sgx}. In practice, only second term which is sourced by EM background field  and GW pass the filter. The frequency of the modulated wave is $|(\omega_\gamma-\omega_g)-\omega_\gamma|=\omega_g$. If we can detect the microwave with this frequency $\omega_g$, we can get an evidence of GW.

Homodyne signal:
We show the schematic picture of the signal from homodyne measuremnt in \cref{fig:schematichomodyne}. For simplicity, we discuss for classical case, but the essence of quantum case is the same \cite{garrison2008quantum}. Let $r,t$ be reflection, transmission coefficient of the splitter, respectively. They obey a constraint
\begin{align}
|r|^2+|t|^2&=1,\\
rt^*+r^*t&=0.
\end{align}
From the second relation, we set $r^*t=i|rt|$. 
The amplitude of the diode 1,2 $E_{D1},E_{D2}$ are
\begin{align}
    E_{D1}&= rE_1+tE_2, \\
    E_{D2}&= tE_1+rE_2.
\end{align}
The difference of output signal is
\begin{align}
    S_{\text{hom}}&= |E_{D2}|^2-|E_{D1}|^2\nonumber\\
    &=(1-2|t|^2)|E_1|^2-(1-2|t|^2)|E_2|^2+4|tr|\Im[E_1^*E_2].
\end{align}
For ideal beam splitter with $|r|^2=|t|^2=1/2$, the signal is
\begin{align}
    S_{\text{hom}}&= 2\Im[E_1^*E_2].
\end{align}
It is possible to obtain the modulation signal originating from the gravitational wave background.

\begin{figure}
    \centering
    \includegraphics[width=10cm]{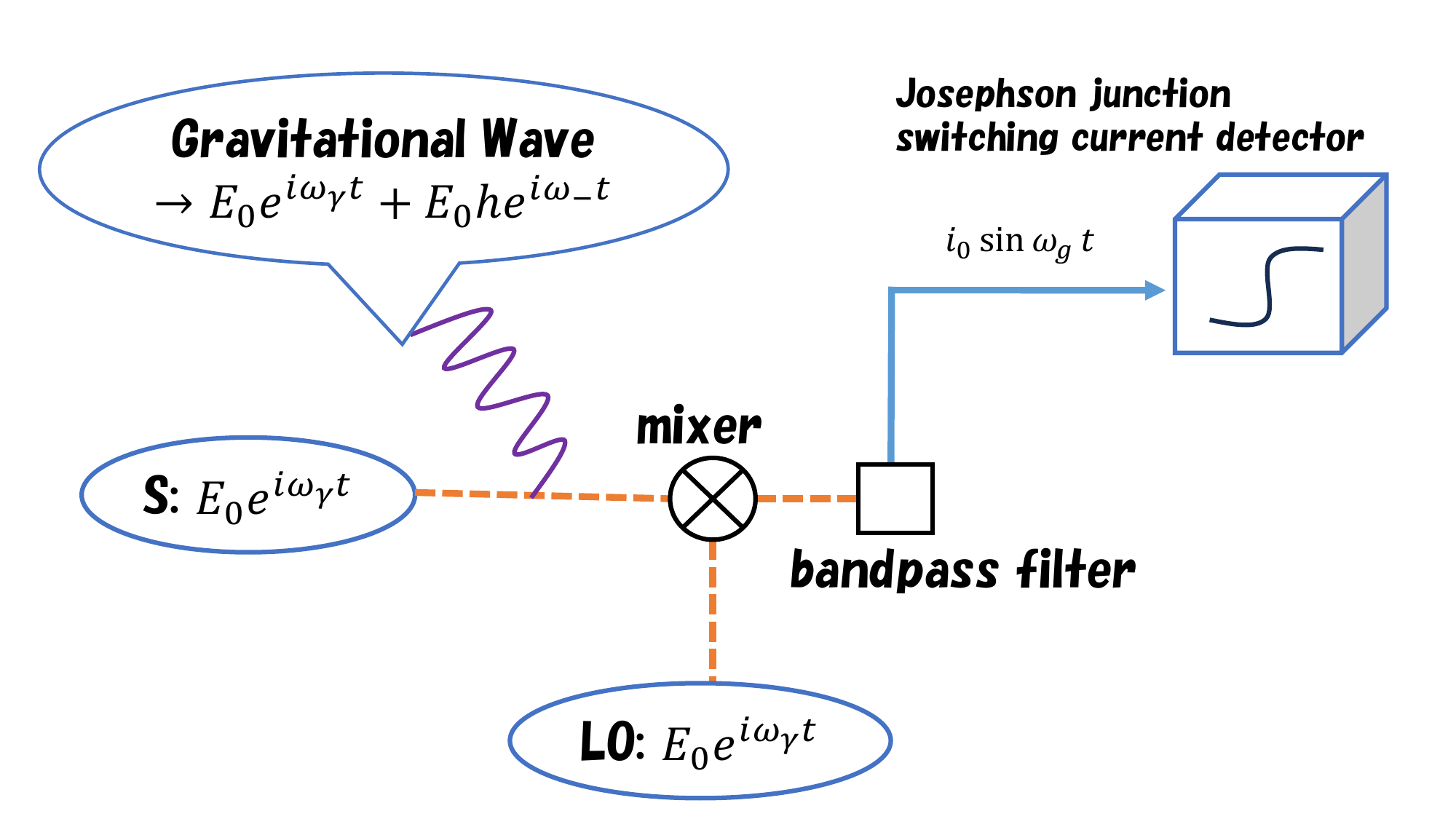}
    \caption{Schematic picture of lock-in technique. By mixing the signal and local oscillator to pass the band pass filter, we can pick up a modulation signal from the mixture of EM wave and gravitational wave. This microwave current can be directly probed by Josephson junction sensor.}
    \label{fig:schhematicmixer}
\end{figure}

Signal from Lock-in technique: Lock-in technique is widely used technique. We show the schematic picture in \cref{fig:schhematicmixer}. By injecting the signal and local oscillator field into mixer to pass the band pass filter, one can pick up a signal with power $E_0^2h$ and frequency $\omega_g$. This signal propagates through the coaxial cable to the Josephson junction. This type of measurement are considered by wide range of people \cite{yates2011photon,de2014fluctuations,zgirski2018nanosecond,wang2018fast,paolucci2020hypersensitive,zhang2024ultra} because it plays an important role in various field from quantum information to astronomy.

\section{Josephson junction Switching current detector (JJSCD)}\label{sec:JJSCD}
\begin{figure}
    \centering
    \includegraphics{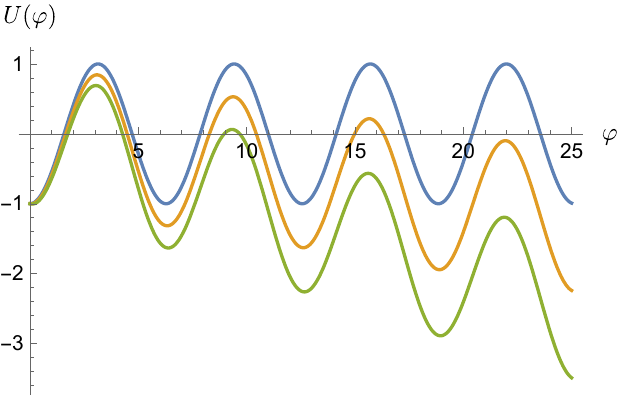}
    \caption{Plot of the Washboard potential $U(\varphi)=E_{J0}(-i\varphi-\cos\varphi)$ in the case of $E_{J0}=1$. The blue curve is for $i=0$, the yellow curve is for $i=0.05$, and the green curve is for $i=0.1$. }
    \label{fig:washboardex}
\end{figure}

Here we describe the Josephson junction (JJ) Switching current detector (SCD) based on \cite{blackburn2020reassessment,krasnov2024resonant}. josephson junction is given by Kirchhhoff law
\begin{align}
    C \left(\frac{\Phi_0}{2\pi}\right)^2\ddot{\varphi}+\left(\frac{\Phi_0}{2\pi}\right)^2\frac{\dot{\varphi}}{R}+\frac{\partial U}{\partial \varphi}=\left(\frac{\Phi_0}{2\pi}\right)(I_b+I_m)
\end{align}
where $C$ is capacitance, $R$ is resistivity, $\Phi_0=2.067\times 10^{-15}$Wb is quantum flux, and $\varphi$ is the phase difference of Josephson junction. In typical cases, the potential is Washboard potential \cite{barone1982physics}
\begin{align}\label{eq:washboard}
    U(\varphi)=E_{J0} (-i\varphi-\cos\varphi)
\end{align}
where $E_{J0}=\left(\frac{\Phi_0}{2\pi}\right)I_0$ is Josephson energy, $i=I/I_0$. The point is that we can regard the equation as particle in metastable potential in one dimension. In \cref{fig:washboardex}, we show the potential for specific values. As you can see, as the current flows, the tilt of the potential increases and current is likely to flow.

Let us see typical values related to the setup. $\omega_{p0}=(2\pi I_{c0}/\Phi_0C)^{\frac{1}{2}}$ is plasma frequency with zero bias. $\Phi_0=2.06\times 10^{-15}$[Wb]  is flux quantum.  For example, take $C=1$pF, $I_{c0}=50\mu$A
\begin{align}
\omega_{p0}
= 389 \text{GHz} \left( \frac{I_{c0} }{50\mu\text{A} } \right)^{\frac{1}{2} } \left( \frac{C}{1\text{pF} } \right)^{-\frac{1}{2} }
\end{align}
We can reach to MHz region if we can use a condensor with capacitance $1$kF.

Dynamics of JJ is described by washboard potential \cref{eq:washboard}. The barrier height is $\Delta U(\varphi)=2E_{J0}(\sqrt{1-i^2}-i\arccos i)$, $i=I/I_{c0}$, $E_{J0}=\Phi_0 I_{c0}/2\pi$ is Josephson energy. 
\begin{align}
E_{J0}
=1.19\times 10^3 \text{K} \left( \frac{I_{c0} }{50\mu \text{A} } \right)
\end{align}

Typical parameter values are $I(t)=I_b \cos\omega_bt$,\ $\omega_b/2\pi=150$Hz,\ $T=1$K, $I_{c0}=50\mu$A. The point is that the bias current is effectively DC because the frequency is low.

It is assumed that the thermal noise dominates and the escape rate is
\begin{align}
\Gamma_0(I) &= a(I) \frac{\omega_p(I) } {2\pi} \exp \left[ -\frac{\Delta U(I)}{k_bT} \right]  \\
a(I)&= \frac{4}{ [ 1+\sqrt{1+Q(I)k_bT/1.8\Delta U(I) } ]^2}
\end{align}

The probability density (escape rate) for switching and the gain are obtained as
\begin{align}
g(I)&= \frac{\Gamma(I) }{dI/dt} [1-G(I)],  \\
G(I)&= \int_0^I dx g(x)dx.
\end{align}
From these two equations, we can obtain the gain as
\begin{align}
G(I) &= 1-\exp \left[ -\int _{0}^{I} dI' \frac{\Gamma(I') }{dI'/dt}  \right].
\end{align}

With microwave, there is an extra current associated with thid microwave $i_{MW}$, and input current is oscillating
\begin{align}
I(t)=I_b \sin\omega_bt+I_{MW} \sin\omega t.
\end{align}
and escape rate is enhanced.
\begin{align}
\Gamma_{MW}=\gamma\Gamma_0
\end{align}
Since this contributes additionary to the bias current, it is expected that switching current is smaller than original value without $i_{MW}$.

For non-resonant case $\omega\ll\omega_p$, the escape rate is obtained by averaging over the period $\tau=2\pi/\omega$
\begin{align}
\Gamma_{MW}&= \frac{1}{\tau} \int_0^\tau \Gamma_0(t)dt.
\end{align}

For resonsnt case,
For high $Q\gg1$, when resonance happens, the enhancement factor is given by the fitting formula \cite{PhysRevB.36.58}
\begin{align}
\log\gamma&\simeq \frac{5E_{J0} \Delta U}{(k_BT)^2} \frac{i_{MW}^2Q}{(\omega_p/\omega_{p0})^2} f(x),\ x=\frac{\omega}{\omega_p}-1 \\
f(x<0)&= Q \left[ \frac{\mathrm{e}^{9x}}{2Q+9} \left(1-2x+\frac{2}{2Q+9} \right) +\frac{\mathrm{e}^{2Qx}-\mathrm{9x}}{9-2Q}\left( 1+\frac{2}{9-2Q}\right)+\frac{2x\mathrm{e}^{9x}}{9-2Q}\right]   \\
f(x>0)&= Q\mathrm{e}^{-2Qx} \left[ \frac{1}{9+2Q}+\frac{2}{(9+2Q)^2}  \right]
\end{align}

\begin{figure}[ht]
\includegraphics[width=8cm] {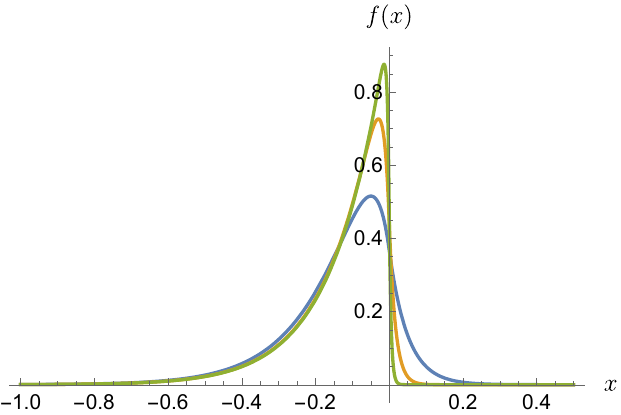} 
\caption{Plot for the examples of $f(x,Q)$. The blue curve is for $Q=10$, the green curve is for $Q=30$, and the yellow curve is for $Q=100$.}
\end{figure}

The point is that this enhancement factor has a sharp cutoff beyond $\omega_p$. From the detector viewpoint, this is more sensitive to lower frequency with respect to resonant frequency $\omega_p$. We can freely tune the $\omega_p$ to hunt the signal originating from GW with frequency $\omega_g$.

\section{Sensitivity estimate}\label{sec:sensitivityestimate}
We make an estimate for the modulated microwave input from the background EM wave and SGWB based on \cite{krasnov2024resonant}. If the induced power is larger than that from thermal noise, we can see the signal. This determines the Noise Equivalent Power (NEP). According to the paper \cite{krasnov2024resonant}, the typical value of NEPs are
\begin{align}
\text{NEP}\simeq
    \begin{cases}
    10^{-19}\text{W}/\sqrt{\text{Hz}}\ \text{for non-resonant case}\\ 
    10^{-21}\text{W}/\sqrt{\text{Hz}}\ \text{for resonant case}\\
    10^{-24}\text{W}/\sqrt{\text{Hz}}\ \text{for best value}
    \end{cases}
\end{align}

As noted in the paper \cite{krasnov2024resonant}, the above two values can be achieved by realistic experimental setup \cite{krasnov2021design,krasnov2023distributed}. Even the best value might be achievable to the level of NEP $\sim10^{-25}$W$/\sqrt{\text{Hz}}$ in the future \cite{paolucci2020hypersensitive}.

In particular, these values are realized experimentally \cite{Lee2019GraphenebasedJJ}. 
We comment that non-resonant sensitivity can be achieved for photo diode \cite{Xie2018SqueezingenhancedHD}, 
also the resonant sensitivity is expected to reach using technology such as squeezing \cite{Feng:2022omd}. Therefore, we expect that the resonant sensitivity is possible in near future and hope that we can use the technology with optimistic level.

We use a input laser With $1$mW as an example \footnote{This value is typical value as large as laser pointer \url{http://www.pmaweb.caltech.edu/~phy003/labs/interferometryhandout5hqM1.pdf}. Note that it might be possible to use $\mathcal{O}(10^2)$W laser \cite{takaku2021large}, or \url{https://lightcon.com/product/carbide-femtosecond-lasers/} in the future or $\mathcal{O}(1)$W laser \cite{yao2023ghz}. If it is possible, we can gain an enhancement from these laser input. I do not know what will happen in the future.}, the Signal to Noise Ratio (SNR) is accordingly
\begin{align}
\text{SNR}\simeq \left(\frac{P_{\text{laser}}}{10^{-3}\text{W}}\right)\cdot
    \begin{cases}
    10^{16}h\sqrt{\text{Hz}}\ \text{for non-resonant case}\\ 
    10^{18}h\sqrt{\text{Hz}}\ \text{for resonant case}\\
    10^{21}h\sqrt{\text{Hz}}\ \text{for best value}
    \end{cases}
\end{align}

\begin{figure}
    \centering
    \includegraphics{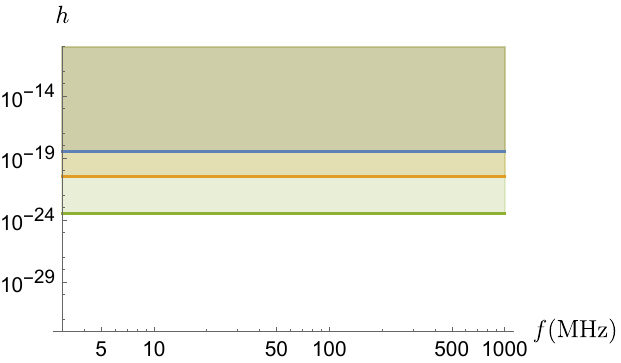}
    \caption{The plot of expected sensitivity for shear $h$. Note that our setup is suited for SGWB. Blue regions are from BBN bound, yellow, green , red regions are from non-resonant, resonant, optimistic cases, respectively.}
    \label{fig:sensitivityflat}
\end{figure}

\begin{figure}
    \centering
    \includegraphics{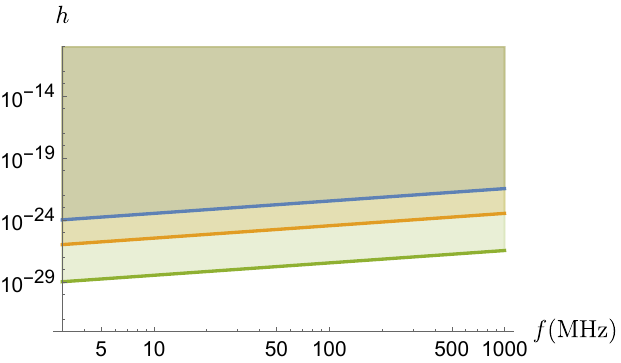}
    \caption{Plot for the expected sensitivity if we can utilize the enhancement factor from the frequency $\omega_\gamma/\omega_g$ with $\omega_\gamma=1$THz laser.}
    \label{fig:sensitivityhp}
\end{figure}

\begin{figure}
    \centering
    \includegraphics{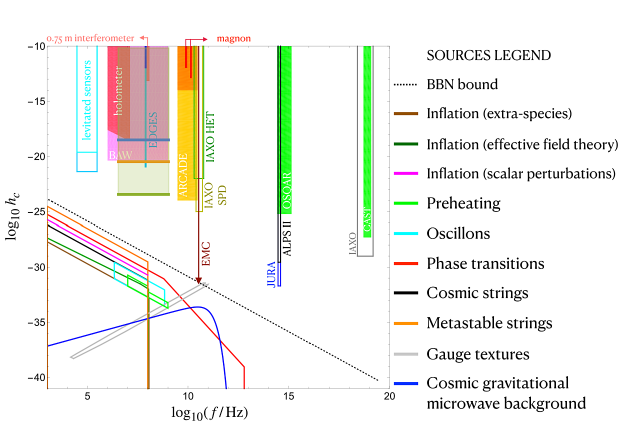}
    \caption{plot of expected sensitivity for shear $h$ in general case. The frequency range is from 3MHz to 1GHz which takes 1day for each frequency by 1MHz. The blue region corresponds to the experimentally realizable region, the yellow region corresponds to the reachable region in the future, and green region corresponds to the optimistic region if we can utilize the ultimate sensitivity. The reference sensitivity compares with \cite{Aggarwal:2020olq}. Note that it will take three years to scan all the regions in this estimate.}
    \label{fig:sensitivity1}
\end{figure}

\begin{figure}
    \centering
    \includegraphics{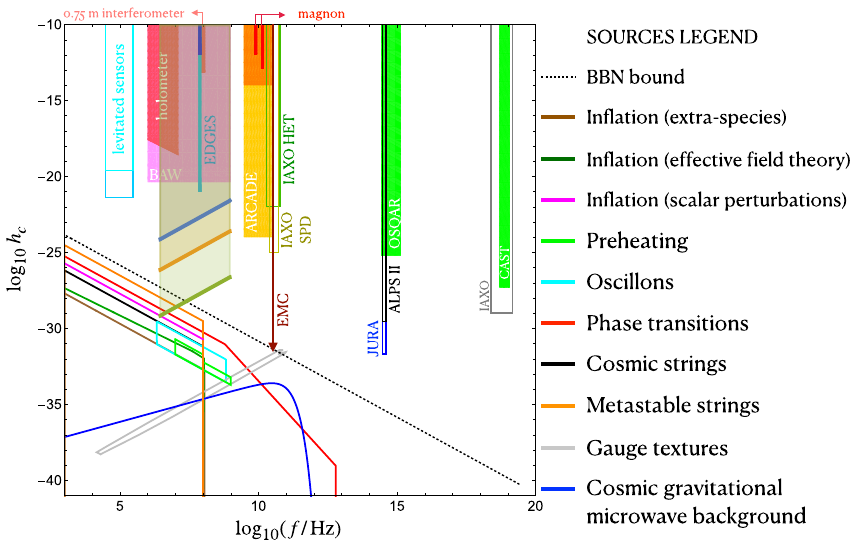}
    \caption{Plot of expected sensitivity for shear $h_+$ with enhancement factor $\omega_\gamma/\omega_g$ with laser frequency $\omega_\gamma=1$THz.  The blue region corresponds to the experimentally realizable region, the yellow region corresponds to the reachable region in the future, and green region corresponds to the optimistic region if we can utilize the ultimate sensitivity. The reference sensitivity compares with \cite{Aggarwal:2020olq}.}
    \label{fig:sensitivity2}
\end{figure}

If we use 1 day for each frequency band
\begin{align}
\text{SNR}\simeq
    \begin{cases}
    3\times 10^{18}h\ \text{for non-resonant case}\\ 
    3\times 10^{20}h\ \text{for resonant case}\\
    3\times 10^{23}h\ \text{for best value}
    \end{cases}
\end{align}
Therefore requiring $\text{SNR}=1$ go to $h\sim3\times 10^{-19},\ 3\times10^{-22},\ 3\times10^{-24}$, respectively. We show these sensitivities in \cref{fig:sensitivityflat,fig:sensitivity1}. It seems that the present estimate does not win the BBN bound. However, it might be possible to beat this bound if we utilize the $\omega_\gamma/\omega_g\sim10^5$ enhancement factor with THz input EM wave in $\mathcal{O}(10)$MHz region. We show the sensitivity including the enhancement factor in \cref{fig:sensitivityhp,fig:sensitivity2}. Although this will be allowed only for $h_+$, it will be fun to explore the region bellow the BBN bound, where we might be able to see the footprint of the early universe region where we cannot see.

\section*{Conclusion}
In this paper, we explored the possibility to detect the signal of stochastic gravitational wave background (SGWB) based on heterodyne (homodyne) measurement with Josephson Junction Switching Current Detector (JJSCD). The conceptual strength of the paper is that the resonant frequency is tunable from MHz to THz. The moderate sensitivity is competitive to other proposals. In the optimistic case, the sensitivity can be close to the BBN bound, furthermore, in the case of high frequency input and low frequency target GW enhancement $\omega_\gamma/\omega_g$ might help to go to the deeper region below the BBN bound. If such case is realized, we might be able to see the footprint of the early universe from the region where we can get access only through the GW. This direction can be one of the ways to search for new physics hidden in early universe for the next generation.

\section*{Acknowledgement}
We appreciate Vladmir Krasnov, Hitoshi Murayama, Elisa Ferreira, Tom Melia, Ippei Obata, Kloian Luzanov, Genta Osaki, Akira Miyazawa, Kosuke Aizawa, Ryosuke Akizawa, and Tomotake Matsumura for conversation, encouragement. This work was supported by JSPS KAKENHI Grant Number 24KJ0613.

\appendix
\section{Photon frequency shift}\label{sec:Photonfrequencyshift}
Let us consider a situation that we emit a photon with frequency $\omega_\gamma$ from the source (S) to the detector (D). We assume that D is on positive x axis. The photon 4 momentum is
\begin{align}
    p^\mu|_{t=0}=(\omega_0,\omega_0,0,0)
\end{align}
The frame with metric $g_{\mu\nu}$, an observer with four-velocity $u^\mu$ will measure a photon frequency $\omega_\gamma =-g_{\mu\nu}p^\mu u^\nu$. We would like to study a tiny effect from GW that passes S and D initially at rest. We can write the metric, four momentum and four velocity as
\begin{align}
    g_{\mu\nu}&= \eta_{\mu\nu}+h_{\mu\nu},\label{eq:BGmetric}\\
    p^\mu&=(\omega_0,\omega_0,0,0)+\delta p^\mu,\\
    u^\mu&=(1,0,0,0)+\delta u^\mu,\\
    \eta&=
    \begin{pmatrix}
        -1&0&0&0\\
        0&1&0&0\\
        0&0&1&0\\
        0&0&0&1
    \end{pmatrix},
\end{align}
where $h_{\mu\nu},\ \delta p^\mu,\ \delta u^\mu$ are $\mathcal{O}(h)$ correction. Here $h$ is the GW amplitude (strain). The inverse metric is
\begin{align}
    g^{\mu\nu}&=\eta_{\mu\nu}-h_{\mu\nu}
\end{align}

We can obtain 
\begin{align}
    \omega_\gamma&=
    -(g_{tt}p^tu^t+g_{tx}p^tu^x+g_{xt}p^xu^t+g_{xx}p^xu^x)\nonumber\\
    &=\omega_0(1+\delta u^t-\delta u^x-h_{tt}-h_{tx})+\mathcal{O}(h^2)
\end{align}
(many terms are in the $\mathcal{O}(h^2)$)

Basically, the photon propagates the vacuum, the momentum $p^0$ obeys the geodesic equation
\begin{align}
    \frac{dp^t}{d\lambda}&= -\Gamma^t_{\mu\nu}p^\mu p^\nu
    =-\omega^2_0(\Gamma^t_{tt}+2\Gamma^t_{xt}+\Gamma^t_{xx})
\end{align}
where 
\begin{align}
    \Gamma^\mu_{\nu\kappa}
    &=\frac{1}{2}g^{\mu\lambda}\left[\partial_\kappa g_{\lambda\nu}+\partial_\nu g_{\lambda\kappa}-\partial_\lambda g_{\nu\kappa}\right] \nonumber\\
    &= \frac{1}{2}\eta^{\mu\lambda}\left[\partial_\kappa h_{\lambda\nu}+\partial_\nu h_{\lambda\kappa}-\partial_\lambda h_{\nu\kappa}\right] 
\end{align}
is a connection for the background metric \cref{eq:BGmetric}.
Specifically,
\begin{align}
    \Gamma^t_{tt}
    &=\frac{1}{2}\eta^{t\lambda}\left[\partial_t h_{\lambda t}+\partial_t h_{\lambda t}-\partial_\lambda h_{tt}\right]
    =-\frac{1}{2}\partial_th_{tt}
\end{align}

\begin{align}
    \Gamma^x_{tt}
    &=\eta^{x\lambda}\left[\partial_t h_{\lambda t}+\partial_t h_{\lambda t}-\partial_\lambda h_{tt}\right]
    =\partial_th_{tx}-\frac{1}{2}\partial_xh_{tt}
\end{align}

\begin{align}
    \Gamma^t_{xt}
    &=\frac{1}{2}\eta^{t\lambda}\left[\partial_t h_{\lambda x}+\partial_x h_{\lambda t}-\partial_\lambda h_{xt}\right]
    =-\frac{1}{2}\partial_x h_{tt}
\end{align}

\begin{align}
    \Gamma^t_{xx}
    &=\frac{1}{2}\eta^{t\lambda}\left[\partial_x h_{\lambda x}+\partial_x h_{\lambda x}-\partial_\lambda h_{xx}\right]
    =-\partial_x h_{tx}+\frac{1}{2}\partial_t h_{xx}
\end{align}

We can think that the light pass the light cone. By substituting the above expressions, we can obtain the master formula for the frequency by integrating $dp^0/d\lambda$
\begin{align}
    \frac{\omega^D_{\gamma}-\omega^S_\gamma}{\omega^D_\gamma}
    &=-\frac{\omega_0}{2}\int_0^{\lambda_D}d\lambda' \partial_t[h_{tt}+2h_{xt}+h_{xx}]_{x^\mu=x^\mu_{\lambda',0}}\nonumber\\
    &+[\delta u^t-\delta u^x](\lambda_D)-[\delta u^t-\delta u^x](\lambda_S).
\end{align}
One way to test GW is to see this frequency shift that is explored in \cite{Bringmann:2023gba}.

\section{Calculation process of the EM wave propagation in gravitational wave background}\label{sec:Caldetail}
In this section, we write about the calculation process in \cref{sec:EMprop}.
Maxwell equation in curved spacetime without external source is \cite{Landau1952ClassicalTO}
\begin{align}
    \nabla_\mu (g^{\mu\alpha}F_{\alpha\beta}g^{\beta\nu})
    =\partial_\mu (\sqrt{-g}g^{\mu\alpha}F_{\alpha\beta}g^{\beta\nu})=0
\end{align}
here,
\begin{align}
g_{\mu\nu}
&=\eta_{\mu\nu}+h_{\mu\nu}.
\end{align}
Since 
   $\det(1+h)= \det\exp(1+h)=\exp(\operatorname{Tr}(1+h))=1+\operatorname{Tr}(h)$,
   \begin{align}
    \sqrt{-g}&=1+\frac{1}{2}h
\end{align}

Then Maxwell equation becomes
\begin{align}
    0&=\partial_\mu \left[ \left(1+\frac{1}{2}h\right)g^{\mu\alpha}F_{\alpha\beta}g^{\beta\nu}  \right] \nonumber\\
    =&\partial_\mu \left[ (\eta^{\mu\alpha}-h^{\mu\alpha})F_{\alpha\beta}(\eta^{\beta\nu}-h^{\beta\nu})  \right]+\frac{1}{2}\partial_\mu [hF^{\mu\nu}]\nonumber\\
    =&\partial_\mu F^{\mu\nu}-\partial_\mu [h^{\mu\alpha}F_\alpha^\nu+F^\mu_\beta h^{\beta\nu}]+\frac{1}{2}\partial_\mu [hF^{\mu\nu}]\nonumber\\
    =&\partial_\mu F^{\mu\nu}+\partial_\mu \left[\frac{1}{2}hF^{\mu\nu}-h^{\mu\alpha}F_\alpha^\nu-F^\mu_\beta h^{\beta\nu}\right]\nonumber\\
    &\equiv \partial_\mu F^{\mu\nu}-j_{\text{eff}}^{\nu}.
\end{align}

We evaluate the left hand side and right hand side to compare each other
We can evaluate the left hand side
\begin{align}
    \partial_\mu\Delta F^{\mu\nu}
    &= -\partial_t^2\Delta A^\nu+\partial_x^2\Delta A^\nu -\partial_x\partial^\nu\Delta A^x
\end{align}
For $\nu=t$
\begin{align}
    \partial_\mu \Delta F^{\mu t}
    &=\partial_x\partial_t \Delta A^x\nonumber\\
    &=(\omega_g-\omega_\gamma)(\omega_gc_\theta-\omega_\gamma)a_-^x\cos_- \nonumber\\
    &+(\omega_g+\omega_\gamma)(\omega_gc_\theta+\omega_\gamma)a_+^x\cos_+
\end{align}
For $\nu=x$
\begin{align}
    \partial_\mu \Delta F^{\mu x}
    &= -\partial_t^2\Delta A^x \nonumber\\
    &=a_x^-(\omega_g-\omega_\gamma)^2\cos_-+a_x^+(\omega_g+\omega_\gamma)^2\cos_+
\end{align}
For $\nu=y$
\begin{align}
    &\partial_\mu\Delta F^{\mu y}
    = (-\partial_t^2+\partial_x^2)\Delta A^y\nonumber\\
    =&-a_y^-[-(\omega_g-\omega_\gamma)^2+(\omega_g c_\theta-\omega_\gamma)^2]\cos_-\nonumber\\
    &-a_y^+[-(\omega_g+\omega_\gamma)^2+(\omega_gc_\theta+\omega_\gamma)^2]\cos_+\nonumber\\
    =&a_y^-[\omega_g^2s_\theta^2-2\omega_g\omega_\gamma(1-c_\theta)]\cos_-+a_y^+[\omega_g^2s_\theta^2+2\omega_g\omega_\gamma(1-c_\theta)]\cos_+
\end{align}
For $\nu=z$
\begin{align}
    &\partial_\mu\Delta F^{\mu z}
    = (-\partial_t^2+\partial_x^2)\Delta A^z\nonumber\\
    =&-a_z^-[-(\omega_g-\omega_\gamma)^2+(\omega_g c_\theta-\omega_\gamma)^2]\cos_-\nonumber\\
    &-a_z^+[-(\omega_g+\omega_\gamma)^2+(\omega_gc_\theta+\omega_\gamma)^2]\cos_+\nonumber\\
    =&a_z^-[\omega_g^2s_\theta^2-2\omega_g\omega_\gamma(1-c_\theta)]\cos_-+a_z^+[\omega_g^2s_\theta^2+2\omega_g\omega_\gamma(1-c_\theta)]\cos_+
\end{align}
In TT gauge, right hand side of \cref{eq:MaxwellOh} is simplified to be
\begin{align}
    &-\partial_\mu \left[\frac{1}{2}hF_0^{\mu\nu}-h^{\mu\alpha}F_{0\alpha}^\nu-F^\mu_{0\beta} h^{\beta\nu}\right] \nonumber\\
    =&h^{\mu\alpha}\partial_\mu F^{\nu}_{0\alpha}+F^{\mu}_{0\beta}\partial_\mu h^{\beta\nu} \nonumber\\
    &=h^{ij}\partial_i(F_0)^\nu_j-(F_0)_{tj}\partial_t h^{j\nu}+(F_0)_{ij}\partial_i h^{j\nu}
\end{align}
The first term vanished because of traceless $h=0$ and Maxwell equation $\partial_\mu (F^0)^{\mu\nu}=0$

For $\nu=t$, the right hand side is
\begin{align}
    h_+s_\theta c_\theta A_0\omega_\gamma^2 \left[\frac{1}{2}\cos_++\frac{1}{2}\cos_-\right]
\end{align}

For $\nu=x$ the right hand side is
\begin{align}
    &h^{13}\partial_x (F_0)_{xz}-(F_0)_{tz}\partial_t h^{zx}+(F_0)_{xz}\partial_xh^{zx} \nonumber\\
    =& -\frac{1}{2}A_0\omega_\gamma h_+s_\theta c_\theta \left[\{\omega_\gamma+\omega_g(1-c_\theta)\}\cos_++\{\omega_\gamma-\omega_g(1-c_\theta)\}\cos_-\right]
\end{align}
For $\nu=y$
\begin{align}
    &h^{ij}\partial_i(F_0)^y_j-(F_0)_{tj}\partial_th^{jy}+(F_0)_{ij}\partial_ih^{jy} \nonumber\\
    =&-\frac{1}{2}A_0h_\times\omega_\gamma \omega_g c_\theta(1-c_\theta)[\cos_+-\cos_-]
\end{align}
For $\nu=z$
\begin{align}
    &h^{ij}\partial_i(F_0)^z_j-(F_0)_{tj}\partial_t h^{jz}+(F_0)_{ij}\partial_i h^{jz}\nonumber\\
    =&\frac{1}{2}A_0h_+\omega_\gamma [s_\theta^2\omega_\gamma (\cos_++\cos_-)+\omega_gc_\theta^2(1-c_\theta)(-\cos_++\cos_-)]\nonumber\\
    =&\frac{1}{2}A_0h_+\omega_\gamma(1-c_\theta)[(\omega_\gamma(1+c_\theta)-c_\theta^2\omega_g)\cos_++(\omega_\gamma(1+c_\theta)+c_\theta^2\omega)\cos_-]
\end{align}
To compare each equation, we can obtain the EM wave amplitude of $\mathcal{O}(h)$. 
\begin{align}
    a^{\pm}_x&= -\frac{1}{4}A_0h_+s_{2\theta}\frac{\omega_\gamma(\omega_\gamma\pm\omega_g(1-c_\theta))}{(\omega_\gamma\pm\omega_g)^2}
    \simeq -\frac{1}{4}A_0h_+s_{2\theta}+\mathcal{O}\left(\frac{\omega_g}{\omega_\gamma}\right),\\
    a^{\pm}_y&= \frac{\pm A_0h_\times\omega_\gamma\omega_g c_\theta}{2\{\omega_g^2(1+c_\theta)\pm 2\omega_g\omega_\gamma\}}
    \simeq \frac{1}{4}A_0h_\times c_\theta+\mathcal{O}\left(\frac{\omega_g}{\omega_\gamma}\right),\\
    a^\pm_z&= \frac{A_0h_+\omega_\gamma\{\omega_\gamma(1+c_\theta)\pm c_\theta^2\omega_g\} }{2\{\omega_g^2(1+c_\theta)\pm 2\omega_g\omega_\gamma\} }
    \simeq \pm\frac{1}{4}A_0h_+\frac{\omega_\gamma}{\omega_g}(1+c_\theta)+\mathcal{O}\left(\frac{\omega_g}{\omega_\gamma}\right).
\end{align}
\printbibliography
\end{document}